\documentclass[12pt]{article}
\usepackage{graphicx,amssymb,amsfonts,amsmath,epsfig}

\makeatletter
\DeclareSymbolFont{AMSb}{U}{msb}{m}{n}
\DeclareSymbolFontAlphabet{\mathbb}{AMSb}
\renewcommand{\section}{\@startsection{section}{1}{\z@}%
                                    {-7ex \@plus -1ex \@minus -.2ex}%
                                    {2.5ex \@plus.2ex}%
                                    {\normalfont\large\scshape\centering}}
\renewcommand{\subsection}{\@startsection{subsection}{2}{\z@}%
                                       {-5ex \@plus -1ex \@minus -.2ex}%
                                       {1.5ex \@plus.2ex}%
                                       {\normalfont\normalsize\scshape}}
\renewcommand{\subsubsection}{\@startsection{subsubsection}{3}{\z@}%
                                       {-5ex \@plus -1ex \@minus -.2ex}%
                                       {1.5ex \@plus.2ex}%
                                       {\normalfont\normalsize\scshape}}

\renewcommand\@seccntformat[1]{\ignorespaces\csname #1name\endcsname\space
                               \csname the#1\endcsname.\quad}   
\def\eqnarray{%
   \stepcounter{equation}%
   \def\@currentlabel{\p@equation\theequation}%
   \global\@eqnswtrue
   \m@th
   \global\@eqcnt\z@
   \tabskip\@centering
   \let\\\@eqncr
   $$\everycr{}\halign to\displaywidth\bgroup
       \hskip\@centering$\displaystyle\tabskip\z@skip{##}$\@eqnsel
      &\global\@eqcnt\@ne$\;\hfil{##}$\hfil
      &\global\@eqcnt\tw@$\;\displaystyle{##}$\hfil\tabskip\@centering
      &\global\@eqcnt\thr@@ \hb@xt@\z@\bgroup\hss##\egroup
         \tabskip\z@skip
      \cr}
\ifcase \@ptsize
\or
\or
\fi
  \divide\@tempdima\baselineskip
  \@tempcnta=\@tempdima
\makeatother
\begin{document}

\newcommand{\temh}{{t_{\rm H}}}
\renewcommand{\thefootnote}{b}

\thispagestyle{empty}

\begin{flushright}\scshape
PUPT-2301
\end{flushright}
\vskip1cm

\begin{center}

{\LARGE\scshape A note on the string spectrum at the Hagedorn temperature
\par}
\vskip15mm

\textsc{J. D. Madrigal$^a$ and P. Talavera\footnote{On leave of absence from FEN, UPC}}
\par\bigskip
{\em
$^a$Departament de F{\'\i}sica i Enginyeria Nuclear,
Universitat Polit\`ecnica de Catalunya,\\
Comte Urgell 187, E-08036 Barcelona, Spain}\\[.1cm]
{\em
$^b$Joseph Henry Laboratories, Princeton University, Princeton, NJ 08544, USA
}\\[.1cm]
\vspace{5mm}
\end{center}

\section*{Abstract}

We discuss semi-classical string configurations at finite temperature. We 
find that those soliton solution in the background describing type IIA strings
disappear or become divergent when we approach the Hagedorn temperature in the strong coupling regime. These findings together with a semi-classical analysis for the Hawking radiation let us to think that Hawking radiation is mainly driven by the existence of highly excited states. As by side, we check that beside the thermodynamical instability the system is dynamical unstable before reaching the Hagedorn temperature.

\vspace{4cm}
\noindent May 2009

\newpage
\setcounter{page}{1}

\setcounter{equation}{0}

\section{Motivation}
\label{sec:intro}

The $T\neq 0$ QCD phase transition plays an important role in the physics of the early universe and of the heavy ion collision. At high temperatures 
the statistical boostrap model \cite{ Hagedorn:1965st} was the first evidence for an
asymptotic dependence of the density of the hadronic states on the mass
\begin{equation}
\label{hag}
\rho(m)\sim f(m) \exp({m\over t_{\rm H}})\,.
\end{equation}
This implies a 
divergent behavior for the thermal partition function above the Hagedorn temperature, $\temh$, and leads originally to think on this temperature as a physically limiting one. Only latter has become clear that the limiting temperature could be interpreted as the phase transition to the partonic degrees of freedom \cite{Kapusta:1981ue}. Nowadays 
lattice results show that the phase transition of the hadronic matter to Quark Gluon Plasma at finite temperature and with a vanishing chemical potential is likely to be a crossover \cite{Aoki:2006we}.

Despite the empirical evidence of the relation (\ref{hag}) up to energies $\approx 2$ GeV \cite{Broniowski:2004yh}, the results are fuzzy at this and even higher energies because a clean identification of masses and decay constants are washed up by the hadrons strong decay. The decay occurs as results of a finite width, and this messed up the experimental values for the mass and width. To overcome this problem, a least theoretically, one can apply the large-N limit of QCD where mesons and glueball widths vanish \cite{'tHooft:1973jz}. In doing so one obtains
a clean evidence that QCD with adjoint fermions in 1+1 dimensions has an Hagedorn spectrum \cite{Kogan:1995nd}. 
This must be also the case for QCD in 3+1 dimensions \cite{Cohen:2009wq} and presumably, even if the theory is not renormalizable,  will hold true in higher dimensions. 

An Hagedorn spectrum was found to arise automatically in string theory, this corroborates the point of view of QCD as giving stringy dynamics for high lying states.
Owing to the AdS/CFT duality \cite{Maldacena:1997re} or its generalizations \cite{Witten:1998zw}, the mentioned first order phase transition occurred in the quantum field theory at the strong coupling limit must have an analogous pattern in the gravity side. This was already envisaged at the level of a weakly interacting gas of strings \cite{Frautschi:1971ij} and latter in a more geometrical fashion  \cite{Susskind:1993if}.
The physical picture that emerged from previous considerations is as follows. If one consider  weakly interacting strings in a finite volume as the fundamental black hole degrees of freedom, at first the strings will predominantly be small. As the Hagedorn temperature is approached long strings begin to form and eventually the configuration will be dominated by a single long string, although presumably remains a small component of small strings. During this process as further energy is pumped into the system most of it is spend on forming long strings rather than increasing the temperature. This behavior is reminiscent of a first-order phase transition with a large latent heat. 

The same behavior was found when one takes the equivalent of the near horizon limit to the NS5 system \cite{LorenteEspin:2007gz}. This is kind of surprising because the near horizon limit  involves the large N$_c$ limit, which one applies to make simplifications on the theory, but in doing so one {\sl assumes that no phase transition is involved in the process}. If this would be the case the theory we end with would presumably belong to a different equivalence class of the initial one. Thus we would have simplified the problem but at the expenses of changing the problem itself. 

While previous studies have been based mostly on ensembles of free strings ours tackles a semi-classical study of strong interacting strings in a curve background. This will allow us to explore some sectors of the spectrum and shade some light on the specific role of $\beta_{\rm Hag}$. This temperature can preclude a phase transitions, as the stringy genus-one corrections suggest  \cite{Kutasov:2000jp}, or can be a limiting one \cite{Aharony:2004xn}. In that respect,
we find evidence of the existence, at least at the semi-classical level, of a second order phase transition, where highly excited stringy modes disappear in favor of low-enegy modes.

This note is laid out as follows: In the following section we describe the essentials of the background. Section \ref{semi}, which constitutes the main body of this note, is devoted to the discussion of different semi-classical string trajectories
in backgrounds describing the evolution of the temperature towards the Hagedorn limit.
We have not exhausted by far all the casuistic, but we have extend the strings in all the possible subspaces of the metric, thus obtaining a wide picture. In section 4 we studied the dynamical stability showing that the system is already dynamically unstable before hitting the Hagedorn temperature. We conclude with a few remarks and implications on our findings.

\section{Basic facts on near-extremal NS5}
\label{clsl}

Our starting point is the low-energy 10 dimensional type IIA string theory action in the Einstein frame
\begin{equation}
\label{action} 
S_{\rm IIA}=\frac{1}{2K_{10}^{2}}\int  d^{10}x  \sqrt{-g}\left(R-\frac{1}{2}g^{\mu\nu}\partial_{\mu}\phi\partial_{\nu}\phi-\frac{1}{12}e^{-\phi}H^2\right) \,,
\end{equation}
where $H$ denotes the NS three-form and $\phi$ is the dilaton field. 

We shall consider an spherical compactification to six dimensions with an $S^3$, and work at the leading $\alpha^\prime$ corrections.  With this proviso 
a solution of (\ref{action}) is given by the following set of fields \cite{Callan:1991at}: 
{\sl i)} A three-form, under which the NS5-branes are magnetically charged, constant along the $S^3$ directions and otherwise vanishing
\begin{equation}
\label{3form}
H_{\alpha\beta\gamma}=- N \epsilon_{\alpha\beta\gamma\omega} \partial^\omega \phi\,,
\quad \alpha,\beta,\gamma,\omega = 6,\ldots, 9\,.
\end{equation} 
{\sl ii)} A dilaton field
\begin{equation}
e^{2\phi}=g_s^2 A(\rho)\,,
\end{equation} 
and {\sl iii)} a metric field, that in the near-extremal case takes the form 
\begin{equation}
\label{metric} 
 ds^2=-f(\rho)dt^2+dx_i dx^i+\frac{A(\rho)}{f(\rho)}d\rho^2+A(\rho)\rho^2d\Omega_{3}^2 \,,
\end{equation}
with $(t,x^i)\,, i=1,\ldots,5,$ the flat directions along the NS5-brane world-volume. The warping factors are given in terms of
the functions
\begin{equation}
\label{warp} 
f(\rho)=1-\frac{\rho_{0}^2}{\rho^2}\;\;\;,\;\;\;A(\rho)=\kappa +\frac{N l_{\rm s}^2}{\rho^2} \,,
\end{equation}
being $\rho_{0}$ the non-extremality parameter and $\kappa$ a discrete constant. 
The value $\kappa=1$ identifies the system of a stack of NS5 branes. 
Eqs. (\ref{warp}) define the two relevant scales in this set up:
The effective length scale  $ \sqrt{N l_s^2}$ and the supersymmetry breaking scale $r_0$.

The near horizon limit of the previous system decouples the mode interactions between the bulk and the brane while a strong interacting theory on the brane remains. This resulting theory is believed to reduce to a string theory without gravity, called Little String Theory \cite{Seiberg:1997zk}. This limit is defined by keeping the string length fixed and taking $g_{\rm s}$ to zero, while at the same time the energy above extremality is also fixed. Explicitly 
\begin{equation}
\label{assumptions}
g_{s}\rightarrow0\;\;,\;\;\rho_{0}\rightarrow 0\;\;,\;\;\mu=\frac{\rho_{0}^2}{g_{s}^2 l_{s}^2}={\rm fixed}\,,
\end{equation}
that corresponds to the choice $\kappa=0$ in (\ref{warp}). The set of expressions (\ref{3form})-(\ref{warp})
fulfill the super-gravity equations of motion {\sl only} for these two choices of the $\kappa$ parameter, $\kappa=0,1$. 
Due to the discreteness of this parameter one can think, in some sense, that the system we study can resemble a two state thermal system: in one of the states we find the lower temperature system, NS5, and in the other with a high-temperature its {\sl near horizon limit}, LST. 
The thermodynamics of this background has entropy as function of energy $S(E) = \beta_\kappa E$, where $\beta_\kappa$ corresponds to the inverse temperature. This is fixed for both models and is given by 
\begin{equation}
\label{temp}
\beta_\kappa=\beta_{\rm Hag}\sqrt{1+\kappa {\rho_0^2\over N}}\,,\quad {\rm with}\quad \beta_{\rm Hag}= 2\pi{\sqrt{N l_s^2}\over m_s}\,.
\end{equation}
In addition
\begin{equation}
\label{evol}
{E\over {\rm Vol}(\Re^5)}= {\rho_0^2\over (2\pi)^5 l_s^6}\,.
\end{equation}
The behavior in (\ref{temp}-\ref{evol}) corresponds to the thermodynamic of a string theory at the Hagedorn temperature, $\beta_{\rm Hag}$ \cite{Maldacena:1996ya}. 
The NS5-brane description of $(2,0)$ Little string theory is valid at very high energies
\begin{equation}
\label{eover}
{E\over {\rm Vol}(\Re^5)} \gg {N\over  l_s^{6}}\,,
\end{equation}
 and in full generality the 
validity of the super-gravity approximation holds if
\begin{equation}
\label{sg}
\rho_0^2\gg N \gg 1\,.
\end{equation}

\section{Semi-classical string spectrum}
\label{semi}
As a first evidence of the existence of a black hole phase transition we explore a part of the spectrum of the model
presented in sec. \ref{clsl} by considering classical string configurations, thereby representing states with large excitations quantum numbers.

In order to obtain the semi-classical trajectories for the string one can start with the Polyakov action in the conformal gauge,
\begin{equation}
I = -{1\over 4\,\pi\alpha^\prime}\int\,d\tau\,d\sigma\,
G_{\mu\nu}\,\partial_\alpha
X^\mu(\tau,\sigma)\, \partial_\beta X^\nu (\tau,\sigma)\,\eta^{\alpha\beta}\,,
\end{equation}
and demand that any configuration fulfills the equation of motion derived from it 
\begin{equation}
 \left( -{\partial G_{\rho \nu}\over
\partial X^\mu}
+2 {\partial G_{\mu\nu}\over\partial X^\rho} \right)\,
\left(\dot X^\rho\, \dot X^\nu -  {X^\rho}'\, { X^\nu}'\right)
+2\,G_{\mu\nu}\,\left( \ddot X^\nu - {X^\nu}'' \right)=0\,,
\end{equation}
where dots and primes denote derivatives with respect to $\tau $ and
$\sigma $, respectively. 
In addition the solution must also satisfy the two Virasoro constraints
\begin{equation}
 G_{\mu\nu}({\dot X}^\mu{\dot X}^\nu+X'^\mu X'^\nu )=0\,,\quad
 G_{\mu\nu}{\dot X}^\mu X'^\nu=0\,.
\end{equation}

{}For a diagonal target metric, as in the case in (\ref{metric}), the energy and the angular momentum 
in a generic angle $\varphi $ are given by
\begin{equation}
E =
- {1\over 2\pi\alpha^\prime}\int _0^{2\pi} d\sigma\ G_{00}(X) \partial_\tau
X^0 \,, \quad
J_\varphi =
{1\over 2\pi\alpha^\prime}\int _0^{2\pi} d\sigma\ G_{\varphi \varphi }(X) \partial_\tau
\varphi \,.
\end{equation}

To set up the calculation we review briefly the simplest of the cases: a rotating string in the Minkowsky part of (\ref{metric}), but probing the rest of the geometry. To describe such string we use cylindrical coordinates for a part of the flat Euclidean space $dx_1^2+dx_2^2=dr^2+r^2 d\beta^2$ and consider the configuration
\begin{equation}
\label{flat}
t=e \tau\,,\quad r(\sigma)\,,\quad\beta=\omega e \tau\,,
\end{equation}
with the rest of coordinates set to constant.
As one realizes immediately from the equation of motion of the transverse coordinate 
this configuration is unstable for the non-extremal metric (\ref{metric}). A similar conclusion, but with a different background, was already reach in \cite{Pons:2003ci}. This is in sharp contrast with the result for the asymptotic flat space, where one makes essentially the replacement $f\to 1$ in (\ref{metric}). {}For this latter case and for the ansatz (\ref{flat}) the action takes the form
\begin{equation}
I=-{1\over 4\pi} d\tau d\sigma (e^2-e^2 \omega^2 r^2+r^{\prime 2})\,,
\end{equation}
where, otherwise states, primes stands for derivatives w.r.t. $\sigma$.
The equation of motion for the radial mode has the well-known solution $r={1\over \omega}\sin(e \omega \sigma)\,,$ and fulfills the energy-momentum relation $E= \omega J_\beta$.

\vspace{1cm}

To continue with
we restrict our attention to configurations that are sensible to the discrete parameter $\kappa$ and checked that for $\kappa=1$ there is indeed a classical solution, but this becomes singular, i.e. unstable,  when we consider the limit $\kappa\to 0$. With this simple exercise we show that both, states characterized by angular momentum quantum numbers, which are conserved, and by winding numbers of the string, which are not conserved because they warp contractible circles, are destabilized when we increase the temperature.

\subsection{String spinning on $S^3$}

We start considering the compact part of the metric and parametrize the $S^3$ sphere by
\begin{equation}
d\Omega_3^2= d\alpha^2+ \sin^2\alpha\, d\theta^2+\cos^2\alpha\, d\varphi^2\,, \quad \alpha\in[0,{\pi\over 2}]\,, \quad \theta\in[0,2\pi]\,,\quad  \varphi\in[0,2 \pi]\,.
\end{equation}
We choose an embedding that mimics the one first considered in \cite{Gubser:2002tv}. The simplest nontrivial state with $S^3$ charge is given by the explicit ansatz 
\begin{equation}
\label{configs3}
t= e  \tau\,,\quad \rho=R\,,\quad \alpha=m \sigma\,,\quad \theta=\varphi= \omega e \tau \,, 
\end{equation}
with the rest of coordinates set to constant. 
The equation of motion reduces to the radial one
\begin{equation}
\kappa (m^2-e^2 \omega^2) R^4 + e^2 \rho_0^2=0\,,
\end{equation}
that shows explicitly that for the case $\kappa=0$ we should not expect an stable solution.
This equation is trivially fulfilled if we locate the string at $R^2 = {e \rho_0\over \sqrt{(e^2 \omega^2-m^2)}}$, provided $\vert e \omega\vert > \vert m \vert$.
Given the set of constants $\{e,\omega,m\}$ the Virasoro constraints imposes the location of the horizon at
\begin{equation}
\rho_0={e^4\omega^2 (1-N\omega^2)+m^2(N m^2-e^2) \over 2 e^3 \omega^2 \sqrt{e^2\omega^2-m^2}}\,.
\end{equation}

The relation between the space-time energy and the angular momenta is linear and given by
\begin{equation}
{E\over \omega J} = 1+{m^2\over e^2\omega^2}\approx 1\,,
\end{equation}
with $J=J_\theta+ J_\varphi$.

\subsection{String spinning on $\Re^6\times S^3$}

The explicit configuration, in the static gauge, we take is the following
\begin{equation}
\label{configr5s3}
t= \tau\,,\quad r=r_0\,,\quad \beta=m(\sigma+\tau)\,,\quad \rho=\rho_0\,,\quad\theta=\omega\tau-n \sigma\,,\quad\varphi=\omega\tau-n\sigma\,, 
\end{equation}
with $ m\,,\omega\,,n >  0$. The rest of coordinates are taken constant. 

The equations of motion amounts to impose a single identity 
\begin{equation}
\label{eomr}
\kappa R^4 ( \omega^2-n^2)= \rho_0^2\,,
\end{equation}
while the two constraint lead to
\begin{eqnarray}
\label{virasoro}
&&m^2 r_0^2-n\omega(N +\kappa R^2)=0\,,\nonumber\\
&&\rho_0^2+R^4\kappa (n^2+\omega^2)+R^2[-1+2 m^2 r_0^2+N(n^2+\omega^2)]=0\,.
\end{eqnarray}
As is evident at first sight from (\ref{eomr}) for $\kappa=0$ the solution is inconsistent and hence the classical configuration (\ref{configr5s3}) is unstable. {F}or $\kappa=1$ there is indeed a stable trajectory that solves both the equation of motion and the Virasoro constraints with 
\begin{equation}
\label{rel1}
R^2={1-N (n+\omega)^2\over 2 \omega(n+\omega)}
\end{equation}
 and with the relation between the parameters
\begin{equation}
\label{rel2}
m^2= {n[1+N (\omega^2-n^2)] \over 2r_0^2(n+\omega)}\,.
\end{equation} 
{}From the positivity of the previous relations one concludes that the string rotates very slowly, at most as  ${\cal O}(n)\sim {\cal O}(\omega)\approx {\cal O}(1/\sqrt{N})$\,.

With this at hand, the Virasoro constraints implies the relation $m r_0^2 J_\beta=n(J_\theta+J_\varphi)$ between the angular momenta associated with the angles $\beta\,,\theta$ and $\varphi$.

Taking into account the positivity of the constraints (\ref{rel1},\ref{rel2}) the relation between the energy and the angular momenta depends crucially on the exact relation between $n$ and $\omega$:
$
E-2 J\sim 1\, \,{\rm for}\,\, n\gg \omega\,\, {\rm and}\,\,   E-2 J \sim {\cal O}(-\sqrt{N})\,\, {\rm for}\,\, n\ll \omega\,\, {\rm or}\,\, n\,\,\approx \omega\,,$ where $J$ refers to the $J_\theta+J_\varphi$ angular momenta.

\subsection{String spinning on $\Re^+ \times S^3$}

Hitherto we have shown that all the configurations corresponding to highly excited states decouple at the Hagedorn transition temperature. We turn now the attention to the
set up which is probably the most interesting one because it was show, in the case of $AdS_5\times S^5$, that twist two operators
\begin{equation}
\label{ttwo}
{\cal O}={\rm Tr}\ \Phi_1({\stackrel{\leftrightarrow}{D}}_+)^l\ \Phi_2\,,
\end{equation}
in the boundary theory corresponded to strings excitation in the bulk. Furthermore, in the case of deep inelastic scattering in QCD, the anomalous dimensions of these operators play an important r\^ole providing the logarithmic dependence on the spin \cite{Gubser:2002tv}.  

The configuration at hand corresponds to a rotating string located in the equator of the $S^3$ 
\begin{equation}
t= e \tau\,,\quad \rho(\sigma)\,,\quad \varphi=e \omega \tau\,.
\end{equation}
Inserting these into the constraints we obtain
\begin{equation}
\label{rs3}
(\rho^\prime)^2 \rho^2 (N+\kappa \rho^2)= e^2 (\rho_0^2-\rho^2)(\rho_0^2+\rho^2(N \omega^2-1)+\kappa \omega^2 \rho^4)\,.
\end{equation}
We notice that: {\sl i)}
The derivative with respect to $\sigma$ of the previous equation, eq. (\ref{rs3}), equals the only non-vanishing eom,  the one for the transverse coordiante.
Thus they are not linear independent, and we discard this latter one. {\sl ii)} Contrary to the previous situations the limit $\kappa\to 0$ is not singular and one can think, in principle, that the string configuration in Little String Theory is smoothly obtained from that of NS5. As we shall see this is not the case. 

Equation (\ref{rs3}) leads to 
\begin{equation}
\label{ss}
d\sigma= {\rho d\rho \sqrt{N+\rho^2 \kappa}\over e \sqrt{(\rho^2-\rho_0^2)(\rho^2(1-N\omega^2)-\rho_0^2-\kappa \rho^4 \omega^2)}}\,,
\end{equation}
from where we can adjust $e$ so that the period of $\sigma$ is $2\pi$.  In the sequel, and to display the differences in some detail, we discuss the two distinct values of $\kappa$ separately.

{\sl i)} {F}or the case $\kappa=0$ there are already some surprises. The two possible turning points dictated by (\ref{rs3}) are
\begin{equation}
\rho_- = \rho_0\,,\quad \rho_+=\rho_0/\sqrt{1-N \omega^2}\,, 
\end{equation}
and in order to have a real value for the latter we need to impose
 \begin{equation}
 1-N\omega^2\ge 0\,.
 \end{equation}
Notice, however, that only $\rho\ge \rho_+$ gives actually a positive value for the cuartic equation derived from the factor inside the squared root of (\ref{ss}). Thus in this situation the string has only one turning point, the lowest, at finite distance from the horizon, while the upper turning point is located at infinity. The only choice we can made is to approach the extremal point at finite distance to the horizon, $\rho_+\to \rho_-$, by lowering the angular velocity of the string $\omega \ll N^{-1/2}$. As we decrease the angular velocity  the strings gets longer and it destabilizes at the critical velocity $\omega_c \sim {\cal O}(N^{-1/2})$ when touches the horizon. As is already evident this kind of situation must  lead to an infinite value for the energy and the angular momentum. In fact one can not deal with infinitely long, rotating strings with a longitudinal finite velocity, and the divergence of these expressions are just signaling this failure. This is in fact a generic 
 condition for preservation of causality in space-time with particle horizons where the strings touches or extends beyond the horizon up to the asymptotic \cite{deVega:1996mv}. In the sequel we check this picture and in addition we see that we can not define a finite quantity from the energy and the angular momenta in this configuration.  
{}For this configuration the energy and the angular momenta are given by
\begin{equation}
\label{Eko}
E= {1\over \pi\alpha^\prime}{\sqrt{N\over 1- N \omega^2}}\int_{\rho_+}^{\infty} {d\rho\over \rho} {\sqrt{\rho^2-\rho_-^2 \over  \rho^2-\rho_+^2}}\,,
\end{equation}
and
\begin{equation}
\label{Jko}
J= {\omega N \over \pi\alpha^\prime} \sqrt{{N \over 1- N \omega^2}}
\int_{\rho_+}^{\infty} d\rho {\rho\over\sqrt{ (\rho^2-\rho_-^2) ( \rho^2-\rho_+^2) }}\,,
\end{equation}
respectively.
To show the divergences clearly we worked out two special cases in those expressions:

The {\bf Long String} limit, accomplished by $N\omega^2\approx 0$, leads to the following result for the previous two expressions
\begin{eqnarray}
E &&= -{1\over 2\pi\alpha^\prime} \sqrt{N}  \int_{0}^{\infty} d\rho {1\over \rho+\rho_0^2} \left(1 + {\rho^2+\rho_0^4\over \rho(\rho+\rho_0^2)} {N\omega^2\over 2}\right)+\ldots\,,\nonumber\\
J&&={1 \over 2 \pi\alpha^\prime} \sqrt{N} \int_{0}^{\infty} d\rho \left({N\omega\over \rho}\right)+\ldots \,.
\end{eqnarray}
Both expressions are UV divergent and in addition the latest one is IR divergent.

The {\bf Short String} limit. To be more concrete,  $\omega = 1/\sqrt{N}+\eta$, where $\eta\to 0$. In this limit the first terms in the expansion give
\begin{eqnarray}
E &&= -{1\over 2\pi\alpha^\prime} {\sqrt{N}\over \rho_0}   \int_{0}^{\infty} d\rho \left\{ {1\over \sqrt{\rho}} +\eta \sqrt{N}\left({\sqrt{ \rho}\over \rho_0^2}+{1\over \sqrt{\rho} } \right)\right\} +\ldots\,,\nonumber\\
J&&={1 \over 2 \pi\alpha^\prime} {N\over \rho_0} \int_{0}^{\infty} d\rho \left({1\over \sqrt{\rho}} + \eta {\sqrt{N }\over \rho_0^2} \sqrt{\rho} \right)+\ldots \,.
\end{eqnarray}
Notice that the leading divergence can be cancelled between the two expressions, $E/J\sim \sqrt{N}$,
but this does not hold for the next order terms.

In conclusion, for the configuration with $\kappa=0$ we end up always with divergent energy and/or angular momenta. This signals, that even the classical trajectory exists,  we can not obtain any useful information from it.

\hspace{0.5cm}

{\sl ii)} 
In the  $\kappa=1$ case appears a new possible extremal point where the string can folds into itself
\begin{equation}
\rho_{\rm m}=\rho_0\,,\quad \rho_\pm^2={1\over 2 \omega^2}\left(1-N\omega^2\pm\sqrt{(1-N\omega^2)^2-4\rho_0^2\omega^2}\right)\,.
\end{equation}
Choosing the extremal points as $\rho_\pm$ the string orbits around the black hole without touching its horizon and the picture matches that of \cite{Armoni:2002xp}. 
Instead if one chooses the lower point of the string as $\rho_{\rm m}$ and the upper either $\rho_-$ or $\rho_+$ the string rotates touching the black hole horizon. This last picture reduces esentially to the $\kappa=0$ case studied previously and as mentioned has several severe drawbacks, thus we shall not consider it further.

In order to have real roots for $\rho_\pm$ we must constraint the possible values of $\omega$ as
\begin{equation}
\label{wc}
\vert\omega N\vert \le \sqrt{N+\rho_0^2}-\rho_0\,.
\end{equation}
{}The energy and the angular momenta can be cast in this case as
\begin{equation}
\label{Ek1}
E= {1\over \pi\alpha^\prime\omega}\int_{\rho_-}^{\rho_+} {d\rho\over \rho} {\sqrt{(\rho^2-\rho_{\rm m}^2)(\rho^2+N) \over  (\rho^2-\rho_-^2)(\rho_+^2-\rho^2)}}\,,
\end{equation}
and
\begin{equation}
\label{Jk1}
J= {1 \over \pi\alpha^\prime}\int_{\rho_-}^{\rho_+} d\rho\, \rho \sqrt{ {(\rho^2+N)^3 \over (\rho^2-\rho_{\rm m}^2)(\rho^2-\rho_-^2)(\rho_+^2-\rho^2) }}\,.
\end{equation}

The {\bf short String} limit is achieved by setting the angular velocity to its critical value (\ref{wc})
$\vert\omega_{\rm c} N\vert \to \sqrt{N+\rho_0^2}-\rho_0$. This shrinks the string to a point particle. In this case, and following the same steps as in the ${\cal N}=4$ SYM case \cite{Armoni:2002xp}, we find for the expressions (\ref{Ek1}-\ref{Jk1})
\begin{equation}
\label{Ek1short}
(2 \alpha^\prime E)^2\rho_0 = (N+\rho_0^2) (\rho_0+\sqrt{N+\rho_0^2} ) \,,
\end{equation}
and
\begin{equation}
\label{Jk1short}
( 2 \alpha^\prime J)^2\rho_0 =  (N+\rho_0^2)\left(4\rho_0^2(\rho_0+\sqrt{N+\rho_0^2})+N(3\rho_0+\sqrt{N+\rho_0^2})\right)\,.
\end{equation}
Relaxing the supergravity conditions (\ref{sg}) the relation between them becomes
\begin{equation}
\label{ej}
{E\over J} = {1\over 2 \rho_0} \left(1 -2 N {\rho_0^2\over J^2} \right)^{1/2}\,,
\end{equation}
that, contrary to the ${\cal N}=4$ SYM case,  is temperature dependent 
\begin{equation}
\label{ejbis}
{E\over J} = {T^2 \over 4 N (T_{\rm Hag}^2-T^2) }- {N\over 2 J^2}\,,
\end{equation}
with $T_{\rm Hag}$ given in (\ref{temp}). Notice that the positivity of (\ref{ejbis}) implies a dependence of the temperature with the angular velocity: vanishing angular momentum corresponds to strings at the Hagedorn temperature, while strings orbiting at high velocity colds down its temperature. Only in the supergravity limit, $\rho^2\gg N$, the relation (\ref{ej}) becomes temperature independent $E/J = 1/(2\rho_0)$.

{\bf Long Strings} are obtained in the limit $\omega\to 0^+$. Then $\rho_-\to \rho_{\rm m}$ and
$\rho_+\to \infty$.
Bearing in mind this approximation we obtain explicitly for the divergent terms in (\ref{Ek1}-\ref{Jk1}) 
\begin{equation}
\label{Ek1long}
E = {1\over 2\pi \alpha^\prime}  \int_0^1{d\rho\over\sqrt{\rho(1-\rho)}} \left( {1\over \eta} + {\cal O}(\eta)\right) \,,\quad
{\rm and}\quad
J =  {1\over 2\pi \alpha^\prime}  \int_0^1{d\rho\sqrt{\rho \over 1-\rho}} \left( {1\over \eta^2} + {\cal O}(\eta^2)\right) \,,
\end{equation}
where $\omega =\eta$ was used as a regulator. From these two expressions follows that $E= \sqrt{J/\alpha^\prime}$. 

\subsection{String spinning on $\Re^+ \times \Re^5$}

As we have seen up to now semi-classical strings configurations that exist on the background of NS5 becomes unstable once we take the near horizon limit. Here a different situation shows up: we find an unstable configuration for the NS5 system and check that this can not lead to a stable one in LST. 

To comfort the procedure outlined in \cite{Kruczenski:2004wg}  we shift to the Nambu-Goto description for the string action
\begin{equation}
\label{nambu}
S = - {1\over 2\pi\alpha^\prime} \int d^2\sigma \sqrt{-\det\vert G_{\mu\nu}\partial_\alpha X^\mu \partial_\beta X^\nu\vert}\,,
\end{equation}
with the target-space $G_{\mu\nu}$ given by (\ref{metric}). We also choose the world-sheet coordinates such that
\begin{equation}
\label{confsp}
X^0 =e  \tau\,,\quad X^1=r= r_0\,,\quad X^2=\beta= e \omega  \tau + m \sigma\,,
\end{equation}
and use the ansatz $X^6=\rho(\sigma)$ for the propagating string. the rest of coordinates are set to constant. These choices leads to the Nambu-Goto action
\begin{equation}
\label{lagg}
S=  - {1\over 2\pi\alpha^\prime} \int d^2\sigma e \sqrt{{m r_0^2 (\rho_0^2-\rho^2)^2 -\rho^{\prime 2} (N+\kappa \rho^2) (\rho_0^2-(1-r_0^2\omega^2)\rho^2)\over \rho^2(\rho^2-\rho_0^2)}}\,.
\end{equation} 

The equations of motion following from this action are satisfied if the radial coordinate is given by
\begin{equation}
\label{sol}
\rho^{\prime}=\pm m {\rho^2-\rho_0^2\over \omega \rho} {1\over\sqrt{N+\kappa \rho^2 }}\,,
\end{equation}
which in turn fixes the folding points, $\rho=\rho_0$ and the infinity. Notice that the fact of dealing with a turning point located at infinity does not allow to construct ``spiky'' configurations at finite angle, $\Delta \theta = \int d\sigma$, as was done in \cite{Kruczenski:2004wg}.

{F}rom the Nambu-Goto action one can compute the energy and the angular momentum to be
\begin{equation}
\label{enNG}
E= {1\over 2 \pi} e \int_{\rho_0}^\infty d\rho {\rho^2(1+\omega r_0^2)-\rho_0^2\over \rho(\rho^2-\rho_0^2)}
\sqrt{N+\kappa \rho^2}\,,
\end{equation}
\begin{equation}
\label{jNG}
J= {1\over 2 \pi}  \omega e r_0^2 \int_{\rho_0}^\infty d\rho  {\rho\over \rho(\rho^2-\rho_0^2)}
\sqrt{N+\kappa \rho^2}\,,
\end{equation}
where we used as an intermediate step eq. (\ref{sol}).

As in the previous configurations we explore the two possible values of $\kappa$.
{\sl i)} Setting $\kappa=0$ in eqs. (\ref{enNG}-\ref{jNG}) leads to
\begin{equation}
E= {1\over 4 \pi} e \sqrt{N} \left( r_0^2\omega \log\left({\Lambda\over \epsilon}\right) + \log\left({\rho_0^2 + \Lambda\over \rho_0^2+\epsilon}\right) \right)\,, \quad
J= {1\over 4 \pi}  \omega e r_0^2 \sqrt{N} \log\left({\Lambda\over \epsilon}\right)\,,
\end{equation}
been $\epsilon$ ($\Lambda$) the IR (UV) regulators respectively.

Notice that none of the two quantities are finite when we remove any of the regulators. Eventhough, we can see whether there is a kind of  ``magnon'' configuration with a finite ratio between both quantities, even if they diverge separately,
\begin{equation}
{E\over J}= \omega+ {\log\left({\rho_0^2 +\epsilon\over \rho_0^2 +\Lambda}\right) \over 
r_0^2 \omega \log\left({\Lambda\over \epsilon}\right)}\,.
\end{equation}
The crucial point to see that this configuration does not exist is no notice that the limits do not commute
\begin{equation}
\label{limits}
\lim_{\Lambda \to \infty} \lim_{ \epsilon\to 0}{E\over J}= \omega+{1\over r_0^2\omega} 
\ne \lim_{\epsilon\to 0} \lim_{\Lambda \to \infty}{E\over J}=\omega\,. 
\end{equation}
One can argue one step further by noticing that when  (\ref{lagg}) is evaluated on the solution (\ref{sol}) it only imposes the constraint $\omega \ne 0$. Thus one think that for large values of $\omega$ both sides of (\ref{limits}) agree. This result must be inconsistent, because as mentioned already an infinite string rotating at any finite angular velocity will violate causality at some of its points.

{\sl ii)} The $\kappa=1$ case contains indeed more involved expressions for the energy and the angular momentum, to which we only give an integral representation
\begin{equation}
E = {1\over 4\pi} e \int_{\epsilon}^\Lambda d\rho \sqrt{N+\rho+\rho_0} \left( {1\over\rho+\rho_0}
+{r_0^2 \omega^2 \over\rho}\right)\,,
\end{equation}
\begin{equation}
J = {1\over 4\pi} e  r_0^2 \omega \int_{\epsilon}^\Lambda {d\rho\over \rho} \sqrt{N+\rho+\rho_0}\,,
\end{equation}
but the essentials reduces to the previous case: nor $E$ neither $J$ have a finite value and their ratio have the very same expression eq. (\ref{limits}) from where we conclude that neither in this case the trajectory is physically relevant.

{F}rom the gauge theory perspective we have deal with operators of the type
\begin{equation}
{\cal O}={\rm Tr}D_+^{l_1} \Phi_1 \ldots D_+^{l_n} \Phi_n\,,
\end{equation}
to be compared with eq. (\ref{ttwo}). Those operators in the boundary theory correspond to a set of point-like particles moving at the speed of light. As we have shown there are no spike configurations and this translates to the absence of point particles.

\section{Lyapunov exponents and dynamical stability}

As we have learned, so far, the spectrum corresponding to excited states in the near-extremal NS5-branes gets unstable at the Hagedorn temperature. This can give the wrong impression that semi-classical trajectories in the NS5-brane system are stable until this temperature.  As we shall show in the sequel, in some simple example, this is not true, thus the system is dynamical unstable long before hitting the thermodynamical instability point.
  
{F}or this purpose we investigate the role of the dynamical stability for the two black holes by calculating the Lyapunov exponents. While the technique represents a useful tool it has uncomfortable shortcomings in he context of general relativity: The exponents are a measure of the deviation of two neighboring trajectories in time and therefore overtly depend on the time coordinate used. The stability analysis of a given orbit is based in the linearization of the equation of motion around this orbit 
\begin{equation} 
\label{lin}
{d\delta X_i\over dt}={\partial H_i(X_j)\over\partial X_j}\Bigg\vert_{X_i(0)} \delta X_j(t)= K_{ij} \delta X_j(t)\,,
\end{equation}
been $K_{ij}$ the linear stability matrix and $X_i$ are phase space coordinates. The solution to the linearized problem (\ref{lin}) can be cast in terms of the evolution matrix as 
\begin{equation}
\delta X_i(t)=L_{ij}(t) \delta X_j(0)\,,\quad {\rm with}\quad \dot{L}_{ij}(t)=K_{ir}L_{rj}(t)\,,\quad {\rm and}\quad L_{ij}(0)=\delta_{ij}\,.
\end{equation}
The eigenvalues of the matrix $L_{ij}$ will leads to the determination of the principal Liapunov exponents at large times
\begin{equation}
\lambda=\lim_{t\to \infty}{1\over t}\log \left({L_{ij}(t)\over L_{ij}(0)}\right)\,.
\end{equation}
{}From the sign of $\lambda$ we shall extract the stability criteria: 
Unstable orbits will have real Lyapunov exponents and will merge as fractals in phase space \cite{Levin:1999zx}.

Following \cite{Chandrasekhar:1985kt} we write the Lagrangian density for the less trivial trajectory we have found, corresponding to a particle localized in $\Re^{(1,5)}$ but probing the transverse coordinate to the brane and for simplicity it will be located at the equatorial plane of the $S^3$ 
\begin{equation}
\label{lag}
2 {\cal L}= - f \left(dt\over ds\right)^2+{A\over f} \left(d\rho\over ds\right)^2+ A\rho^2\left({d\theta\over ds}\right)^2\,,
\end{equation}
$s$ refers to the affine parameter.  From (\ref{lag}) we can define the conjugated variables 
\begin{equation}
\label{mom2}
-\pi_t=E=f {dt\over ds}\,,\quad \pi_\theta=L=A\rho^2{d\theta\over ds}\,,\quad \pi_\rho={A\over f}{d\rho\over ds}\,,
\end{equation}
been the two formers constants of motion.
In what follows we shall refer derivatives to the time $t$ by an overdot and it will be defined through the first expression in (\ref{mom2}). The equations of motion, $d\pi_q/d\tau=\delta{\cal L}/\delta r$, reduce to a two dimensional system
\begin{eqnarray}
\dot{\rho}&&= {(\rho^2-\rho_0^2)^2\over \rho^2(N+\kappa \rho^2)}{\pi_\rho\over E}\,,
\nonumber\\
\dot{\pi}_\rho&&=-{\rho_0^2 E \over\rho(\rho^2-\rho_0^2)} - {(\rho^2-\rho_0^2)(N+\kappa\rho_0^2)\over \rho(N+\kappa\rho^2)^2}{\pi_\rho\over E}+ \kappa {(\rho^2-\rho_0^2)\over \rho(N+\kappa\rho^2)^2}{L^2\over E}\,.
\end{eqnarray} 
Paying only attention to circular orbits the linear stability matrix in (\ref{lin}) gives
\begin{equation}
\label{k}
K_{ij}=
\left(\begin{array}{cc}
0& {(3\rho^2-\rho_0^2)\rho_0^2\over (\rho^2-\rho_0^2)^2\rho^2}E+\kappa {N(\rho^2+\rho_0^2)-\kappa (3\rho^2-5\rho_0^2)\rho^2\over \rho^2(N+\kappa\rho^2)^3} {L^2\over E}\\
{(\rho^2-\rho_0^2)^2\over \rho^2(N+\kappa\rho^2)}{1\over E} & 0\\
\end{array}\right)\,.
\end{equation}
In the eventual case of taking $\kappa=0$ the eigenvalues of (\ref{k}) are
\begin{equation}
\label{eigen0}
\lambda^2 = {\rho^2_0\over N \rho^4}  (3\rho^2-\rho_0^2)\,,
\end{equation}
which are always positive defined.
In the case of $\kappa=1$ the analytical structure of the eigenvalues of (\ref{k}) is a not very enlighten rational function in terms of the impact parameter $D={L\over E}$. {}For the case of circular orbits there is a well established procedure to obtain the impact parameter \cite{Bardeen:1972fi}, that for the particular geometry (\ref{metric}) gives
\begin{equation}
\label{d}
D=  {\rho_0(N+ \rho^2)\over  \rho_0^2-\rho^2}\,.
\end{equation}
Substituting this expression into the eigenvalues one obtains
\begin{equation}
\label{eigen}
\lambda_{\pm}=\pm 2 {\rho_0\sqrt{N+ \rho_0^2}\over \rho(N+\rho^2)}\,.
\end{equation}
Conservation of energy ensures that in our canonical coordinates these eigenvalues appear in pairs, plus minus sign, to conserve the volume of phase space. 
The most relevant feature of (\ref{eigen0}) and (\ref{eigen}) is that they are {\sl real} implying that both black holes have some dynamical instability.

To assess the relevance of the instability we compare the Lyapunov timescale, $T_\lambda=1/\lambda$, to the gravitational time scale, $T_\omega=2\pi/\dot{\theta}$
\begin{equation}
\label{times}
{T_\lambda\over T_\omega} = {1\over 4 \pi} {\rho A(\rho)\over \rho_0 A(\rho_0)}\,.
\end{equation}
The interesting case is when the Lyapunov timescale is too much shorter than the gravitational wave timescale, then the instability is observationally relevant. This is only achieved for
the case $\kappa\to 1$ and trajectories relatively near the horizon, $\rho_0\leq \rho\lessapprox 4 \pi \rho_0$, otherwise the effect is washed out.

\section{Discussion}

There has been constantly a significant effort to the study of the high-temperature behavior of strings. One reason for that is the hope that a better understanding of string properties under extreme conditions will shed some light on the fundamental nature of strings and on the space-time.
In this line, there has been a variety of studies, most of the treatments are based in a gas of weakly interacting strings. Here we changed the point of view and considered strong interacting strings.

Our study precludes, at least, a first phase transition at $\beta_{\rm Hag}$, where none of the thermodynamical quantities studied in \cite{LorenteEspin:2007gz} display any discontinuity. During this transition the strongly interacting strings at high energy ``melt'' to give rise to a stringy gas, that so far has been assumed to interact weakly. Thermal fluctuations in this primordial gas of windings strings could expand after all the three macroscopic spatial dimensions \cite{Brandenberger:1988aj}.
{F}rom the corresponding field theory context our findings can be interpreted as the absence in the spectrum, at high temperatures and energies, 
of single trace operators corresponding to states with large quantum numbers. 
Putting these results together with those in \cite{LorenteEspin:2007gz} we can conclude that, semi-classically, Hawking radiation can be associated with the existence of states with large quantum numbers: when approaching the Hagedorn temperature the emission from the black hole is non-thermal and field theory states with large quantum numbers exists. Once we have reached the phase transition point the emission is pure thermal and the spectrum reduces to less excited states.
Between this two stages, anddu
once the high energy strings have disappear, one can consider low-energy stringy modes that do not fulfill the requirement (\ref{sg}) and can not be treated within our approach. 
At such low-energies, much smaller than the $S^3$ radii, to be compare with (\ref{eover}),  
\begin{equation} 
E < {1\over R_{S^3}}=  {1\over \rho^2+Nl_s^2}\,,
\end{equation}
we can consider that all the directions are effectively non-compact and
the string system can be treated effectively as an ensemble of massless particles in 10 dimensions \cite{Bowick:1989us}. One can attempt to increase the temperature of this multi-string system by adding energy to the system, this lead to the excitation of the oscillations modes of the string and to the formation of long strings.

\vskip 6mm

{\it{\bf Acknowledgments}}

The work of J.\ D.\ M.\ is supported by a MEC short collaboration fellowship.
The work of P.\ T.\ is
partially supported by FPA2007-66665, Consolider CPAN grant SCD2007-00042, Grup Consolidat 2005SGR00564 and by a MEC PR2008-0332 grant.

\bibliography{none}
\bibliographystyle{ssg}
\nocite{*}    

\end{document}